\documentclass[10pt,conference]{IEEEtran}

\usepackage{graphicx}
 \usepackage{color}	
 \usepackage{subcaption}
\usepackage{siunitx}

\usepackage[usenames,dvipsnames]{xcolor}
\usepackage{tcolorbox}
\usepackage{tabularx}
\usepackage{array}
\usepackage{colortbl}
\tcbuselibrary{skins}
\usepackage{amsmath}
\usepackage{tikz-cd}
\usepackage{stmaryrd}

\ifCLASSINFOpdf
\else
\fi
\begin{document}

\title{BANZKP: a Secure Authentication Scheme Using Zero Knowledge Proof for WBANs}

\author{
\IEEEauthorblockN{Nesrine KHERNANE}
\IEEEauthorblockA{Sorbonne Universities \\UPMC Paris 06\\
CNRS LIP6 UMR 7606\\
KhernaneNesrine@gmail.com}
\and
\IEEEauthorblockN{MARIA POTOP-BUTUCARU}
\IEEEauthorblockA{Sorbonne Universities \\UPMC Paris 06\\
CNRS LIP6 UMR 7606\\
maria.potop-butucaru@lip6.fr}
\and 
\IEEEauthorblockN{Claude CHAUDET}
\IEEEauthorblockA{LTCI, CNRS UMR 5141\\
Télécom ParisTech\\
Université Paris-Saclay\\
claude.chaudet@telecom-paristech.fr}
}

\maketitle

\begin{abstract}
Wireless body area network(WBAN) has shown great potential in improving healthcare quality not only for patients but also for medical staff. 
However, security and privacy are still an important  issue in WBANs  especially in multi-hop architectures. 
In this paper, we propose and present the design and the evaluation of a secure lightweight and energy efficient authentication scheme BANZKP based on an efficient cryptographic protocol, Zero Knowledge Proof (ZKP) and a commitment scheme. ZKP is used to confirm the identify of the sensor nodes, with small computational requirement, which is favorable for body sensors given their limited resources, while the commitment scheme is used to deal with replay attacks and hence the injection attacks by committing a message and revealing the key later. Our scheme reduces the memory requirement by 56.13\,\% compared to TinyZKP~\cite{2}, the comparable alternative so far for Body Area Networks, and uses 10\,\% less energy.

\end{abstract}


%
\IEEEpeerreviewmaketitle

\section{Introduction}
\label{sec:relatedwork}


Wireless body area network is a promising technology for various applications, and it shall be increasingly necessary for monitoring, diagnosing and treating populations. Recent medical reports predict that the number of people using home health technologies will reach the 78 million consumers by 2020 instead of 14.3 million consumers in 2014. Body sensors shipments will hit 3.1 million units every year. To address the increasing use of sensors in this area, a new technology called WBAN (Wireless Body Area Networks) has emerged in response to the various disadvantages associated with wired sensors commonly used  to monitor patients in hospitals and emergency rooms. The mess of wires attached to a patient is not only uncomfortable for patients, leading to a very limited mobility and making patients anxious, but it is also difficult to manage for staff. Voluntary disconnections of sensors by patients are very common and reintegrating these sensors properly is difficult if not impossible.

WBANs could hence represent a true advance in digital patient care. However, their characteristics such as the use of a wireless medium with a low SNR, or the multi-hop communication, expose information to multiple types of security and privacy attacks (e.g., eavesdropping, modification, loss, injection), and make these attacks even more likely than in traditional wireless sensor networks. Two classical bricks are classically used to prevent such attacks: nodes authentication and communication encryption. However, their implementation in WBANs is a real challenge. 



Existing security mechanisms, such as asymmetric cryptography, used in wireless networks are inappropriate given the body sensors limitations in terms of power, memory capacity, communication and computational capabilities.

To establish a trust relationship among the WBAN sensors, and to ensure a secure forwarding of collected data from the different nodes of the network to a collection point, an lightweight authentication mechanism must be implemented. 


The primary focus of TinySec~\cite{6}, a popular secure link-layer protocol,  is to ensure a secure communication between sensor nodes. Its designed to be easy to use, to consume little energy and to require a minimal amount of memory. Unfortunately, there is no restriction on keying method, and a single key pair is selected for the whole network which allow an adversary to pollute an entire network by compromising only one single node ~\cite{7}.

To deal with this problem, Luk et al. proposed an efficient solution in MiniSec ~\cite{8}, in which each pair of nodes shares two secret keys, one for each direction of communication. An internal counter for each direction is used as a nonce and incremented at each use of the associated key. The counters must be synchronized on both sides and only the last bits are included in the packet to minimise the transmission energy. The drawback is that every node should keep a counter for each of its neighbours, which are possible senders, resulting in high memory overhead and making the resynchronization of counters a very expensive operation, since the counter can be unsynchronized.

The basic idea behind $\mu$Tesla~\cite{9} is to solve some difficulties of standard Tesla in sensor networks to achieving asymmetric cryptography via delayed disclosure of the symmetric keys. A sender signs messages using the commitment scheme and broadcast the message without disclosing the key. A short time later, the sender broadcasts the key that will not be used in the future. Time synchronization is necessary between the involved nodes~\cite{10}, which increases authentication delay~\cite{2}.

Even if the commitment scheme used in $\mu$Tesla requires approximately 1000 times less computational resources than ECDSA~\cite{11}; the number of packet that should be stored in each node until the disclosure of the keys may require large memory, since the key disclosure is independent from the packets broadcast, and is tied to time intervals.


The TinyPK scheme described in~\cite{13} based on the use of the public key cryptography using RSA with a public exponent equal to 3, and different Diffie Hellman key exchanges to ensure the authenticity of the sink. However, this process increases authentication delay and the  evaluation of the scheme shows that the nodes spend much time realizing public and private key operations. In addition, Das et al. found a vulnerability against masquerade attack of TinyPK in~\cite{14}. Compared to the  RSA crytpo-system, systems based on elliptic curve digital signature algorithm (ECDSA) described in~\cite{15}, are more efficient since they are capable to maintain the same security level with shorter key sizes. However the transmission and verification of public key certificates require an additional power consumption and memory.

Li et al. proposed a secure sensor association and key management scheme for WBAN, called group device pairing (GDP)~\cite{16}, by using an out of band authentication technique. They assume the existence of auxiliary channels and require the users to visually inspect simultaneous LED blinking patterns in order to achieve a good level of authentication. Such human aided verification may not be intuitive to use, and it is unlikely to be appropriate for emergency scenario~\cite{4}.

A distributed prediction based secure and reliable renting framework (PSR) was proposed in~\cite{17} for wireless body area networks. Each node maintains a matrix, in which it stores the link quality measurements between itself and all other nodes in the network during the last $p$ past time slots. They also proposed an authentication scheme that requires computational resources and hence an additional energy consumption.

To cope with these constraints Goldwasser et al.~\cite{1} developed an efficient cryptographic protocol (Zero Knowledge Proof) with small computational requirement and less energy consumption. ZKP can be used in both exchange keys, and authentication mechanisms.

To our best knowledge, the first to use  Zero Knowledge Protocol in WBAN was~\cite{2}. This scheme, called TinyZKP, allows a receiver $R$ to verify that a piece of data originates from a sender $S$ without leaking any secret information. The results in ~\cite{2} demonstrate that the performance of TinyZKP is better compared to other existing approaches (i,e. T-ECDSA, W-ECDSA), in terms of execution time, memory requirement and energy consumption. However TinyZKP used a large pre-distributed set of keys, 20 private keys, and 20 public keys for each node. It requires memory in the nodes and complicates the registration phase. The service provider has to register the public keys of every sensor node (e.g. 120 public keys for 6 nodes) into to the base station (sink).  Furthermore to sign a message TinyZKP used ECDSA algorithm~\cite{11} in which the shortest possible signature size is 320 bits, which requires computational resources.

\paragraph*{Our contribution}

In this paper, we present and prove correctness of BANZKP, a novel ZKP-based solution. It allows two entities to verify their mutual identities with the low computational requirement of a local Zero Knowledge Proof scheme. Zero Knowledge Proof schemes, when used alone, are vulnerable to replay attack~\cite{1}, which can permit an adversary to inject false data once he successfully performed a replay attack. To cope with this problem, BANZKP uses another cryptographic tool:  a Commitment Scheme that allows one party to commit the message and reveal the secret later. The security and efficiency performance of our scheme are evaluated in the OMNET++ simulator, by implementing BANZKP as an add-on to the convergecast routing protocol.
Compared to TinyZKP~\cite{2}, BANZKP reduces the memory requirement by 56,13\% and the energy by 10\%. 

Section \ref{sec:toolbox} discuss several security and privacy issues for WBAN. In Section \ref{sec:ourBANZKP} we describe our authentification scheme. In Section \ref{sec:Security_and_efficiency_Analysis} we analyse its privacy, security and efficiency and compare it to TinyZKP.

\section{Toolbox}
\label{sec:toolbox}

The main goal of our work is to ensure a trust relationship among the WBAN nodes, and ensure a secure and privacy-protecting forwarding process of the medical data collected  by the sensor nodes to the sink. This solution shall take into account the nodes constraints in terms of energy and computation. The secure term can indeed cover many security features, such as data confidentiality, 	
authentication, 
data integrity,
data freshness,
secure management,
availability,
dependability
revocability,
accountability,
or non-repudiation~\cite{22,24,23}.
In BANZKP, we focus on the three main properties: data confidentiality, data authenticity, and data integrity, as most other properties derive from these ones. 

Concerning privacy, Li et al. ~\cite{21} outlined a good and explicit taxonomy of privacy in traditional WSN (that can be heavily borrowed in WBAN), by dividing it into two principal axes: Data-oriented privacy and Context-oriented privacy. Data-oriented privacy concerns the data created or transmitted within the network, while context-oriented privacy cover contextual information such as the location of a node/network, or the timing of traffic flows. In BANZKP, we first focus on data privacy by ensuring that in the case of multihop communication, only the emitter and the sink are able to have access to the unencrypted patient-related data.

To this extent, BANZKP combines two cryptographic tools: a Zero Knowledge Proof scheme and a Commitment scheme that are described hereafter.

\paragraph*{Zero Knowledge Proof (ZKP)}
The main objective of  ZKP schemes is to let two parties, a sender and a verifier, verify the identity of their peer. Both nodes exchange a few challenge/response messages without disclosing any information about a shared secret to the other party and henceforth to any eavesdropper. 
\cite{25} proved that ZKP schemes have the following 4 main properties:
a) the verifier cannot guess any information from the exchanged messages during the challenge/response phase; 
b) the sender cannot cheat the verifier; 
c) the verifier cannot cheat the sender; 
d) the verifier cannot cheat another party by pretending to be the sender.


\paragraph*{Comitment Scheme}
Commitment schemes~\cite{16}, are cryptographic primitives used to prevent eavesdropping by letting a sender transmit an encrypted message to a receiver which does not possess the decryption key yet. The key shall be transmitted later, when the sender receives a signal from the receiver. If used with classical additional techniques, it has the following properties.
a) a receiver cannot cheat and replay the message or use it to make its own calculation;
b) the sender cannot cheat by changing the message after committing it. 

\section{BANZKP Authentication scheme}
\label{sec:ourBANZKP}

BANZKP uses symmetric cryptography to provide data confidentiality, as asymmetric key cryptography requires a high computationally and energy resources, which is not favorable for resources limitation of body sensor nodes. Besides, BANZKP uses the challenge-response mechanism of a ZKP protocol as well as a Commitment Scheme to let the sensors authenticate the sink node. 

BANZKP supposes that a relaying protocol provides and updates valid routes between each node and the data sink. For evaluation, we used the convergecast routing protocol provided by Omnet++, which works in two simple and generic phases. First, to establish the routes the sink broadcasts a Route-Flood message to every node in the network. This message is used by each node to choose a parent towards the sink and build a collection tree. The metric to compare routes can be any additive metric and nodes only maintain a single path towards the sink that will be used in the data transmission phase. Nodes do not know each other and cannot communicate together directly. 




\begin{table}[h]
\caption{Main notations} 
\centering 
\begin{tabular}{c c} 
\hline\hline 
Notation&\multicolumn{1}{c}{Description } \\ [1ex]
\hline 
$ID_{i}$ & \begin{minipage}[t]{0.7\linewidth} The node ID of sensor node i \end{minipage}\\ [1ex]
$K_{x,y}$ & \begin{minipage}[t]{0.7\linewidth}The symmetric session key between x and y\end{minipage} \\ [3ex]
$K_{CS}$ & \begin{minipage}[t]{0.7\linewidth}The commitment scheme key \end{minipage} \\ [1ex]
$V_{0,n}$  & \begin{minipage}[t]{0.7\linewidth}The secret information shared between the sink (node 0) and n \end{minipage} \\ [3ex]
$p_{n,0}$  & \begin{minipage}[t]{0.7\linewidth}The random value chooses by n  \end{minipage} \\ [1ex]
$q_{0,n}$  & \begin{minipage}[t]{0.7\linewidth}The random value chooses by the node 0 \end{minipage}\\ [1ex]
E($K$[M])  & \begin{minipage}[t]{0.7\linewidth}Encryption message M with the session key K \end{minipage}\\[3ex]
RI              & \begin{minipage}[t]{0.7\linewidth}Random interval \end{minipage}\\[1ex]
L(X)            & \begin{minipage}[t]{0.7\linewidth}Length of X \end{minipage}\\[1ex]
\textbar\textbar             & \begin{minipage}[t]{0.7\linewidth}Concatenation operator \end{minipage}\\[3ex]

\hline 
\end{tabular}
\label{tab:hresult}
\end{table}


\subsection{System Parameters and Assumptions}

We consider a network composed by 7 nodes, numbered from 0 to 6, deployed around, on, or implanted into the human body. BANZKP makes  the following assumptions,  which are the same as TinyZKP:
\begin{enumerate}
\item The nodes and the sink are assumed to be protected from physical compromission and trustworthy. This assumption is reasonable because the different nodes and the sink are handled by a patient and can be protected in secure location. Besides, the nodes can be equipped with anti-tampering mechanisms. Therefore we can limit protection to external attacks only.
\item Due to the constrained resources of the body sensor nodes, computationally expensive and energy intensive operations shall be avoided to calculate and transmit keys. Therefore, the different keys and parameters used by BANZKP should be uploaded by an operator in the nodes before deployment.
\item To register a new node as a member of a given WBAN network, or to replace a node that does not work anymore, the sink must be accessible by the operator in order to register the new node, i.e. to upload in the sink shared parameters specific to this node. The use of close-range pairing mechanisms could be used at this stage. 
\end{enumerate}

Under the previous assumptions, for each node $n$, BANZKP uses and maintains the following values: 
\begin{enumerate}
\item $n$ shares with the sink (node 0) a session key $K_{0, n}$, n=\{1, ..., 6\}. The values of $K_{0, n}$ are different for each node, uploaded manually at the node registration phase and should kept secret.
\item $n$ shares with a sink a number $V_{0, n}$, n=\{1, ..., 6\} used for authentication. The values of $V_{0, n}$ are different for each node, , uploaded manually at the node registration phase and should kept secret.
\item $n$ chooses a random number $p_{n, 0}$, n=\{1, ..., 6\} used for authentication with the sink node.
\item The sink chooses randomly one different random number for each node $n$ : $q_{0, n}$, n=\{1, ..., 6\}.
\end{enumerate}

\subsection{BANZKP protocol}
BANZKP is composed of two phases: a registration phase in which an operator physically pairs the nodes and the sink and an online authentification phase, both described below.

\paragraph*{\bf{Registration Phase}}
In this phase, an operator (aka service provider) registers each node with the sink by uploading each secret number \{{$V_{0, 1}$, $V_{0, 2}$,..., $V_{0, 6}$}\} into the sink which is considered as the authentication center, as well as the different shared keys, \{{$K_{0, 1}$, $K_{0, 2}$,..., $K_{0, 6}$}\}. These keys, shared by the sink and each  node, allow sensors to communicate with the sink and ensure a secure data forwarding. 

\paragraph*{\bf{Authentication Phase}}
We suppose that the sensors are deployed at designated places (on/in/around the human body), and that system initialization is finished. When a node $N$ has data to send, it starts the authentication mechanism. Authentication is mutual, which means that the node shall prove its identity to the sink and verify that the sink is the expected one. Our approach is based on the strength of the zero knowledge proof algorithm, and the communication between the sensor node \emph{N} and the sink \emph{0} can be decomposed in the five following steps:





\begin{enumerate}
\item \emph{Sensor node} $\rightarrow$  \emph{sink}: $\displaystyle E\left( K_{0,N} \left[ ID_{N}  ||  V^{p_{N,0}}_{0,N} \right] \right)$ \\

The node $N$ draws $p_{N,0}$, calculates $V^{p_{N,0}}_{0,N}$, concatenates it to its identifier $ID_{N}$, encrypts it with its session key $K_{0,N}$ and sends the entire resulting message to the the sink.

\item \emph{sink} $\rightarrow$ \emph{sensor node}:\\ $\displaystyle E\left( K_{0,N} \left[ ID_{0}  ||  V^{q_{0,N}}_{0,N} || RI \right] \right), E\left( K_{CS} \left[ \left( V^{p_{N,0}}_{0,N} \right)^{q_{0,N}} \right] \right)$ \\

   Upon receiption of the initial message, the sink decrypts it and then proceeds its calculations; it firstly calculates $V^{q_{0,N}}_{0,N}$ and encrypts it with the session key $K_{0,N}$, and then calculates $(V^{p_{N,0}}_{0,N})^{q_{0,N}}$, which has minimum size of 1096 bits, chooses a random interval such as the size of this latter must be 200 bits, and encrypts it with the commitment scheme key $K_{CS}$ (chosen randomly). The beginning of the interval \emph{RI} is encrypted with the session key $K_{0,N}$.

   The encrypted message, which includes the identifier, $ID_{0}$, $V^{q_{N,0}}_{0,N}$, RI and $(V^{p_{N,0}}_{0,N})^{q_{0,N}}$ interval value, is sent to the sensor node N.

\item \emph{Sensor node} $\rightarrow$ \emph{sink}: $\displaystyle E\left( K_{0,N} \left[ ID_{N}  ||  \left( V^{q_{0,N}}_{0,N} \right)^{p_{N,0}} \right] \right)$

    When it receives the message from the sink, the sensor node \emph{N} stores the received commitment message as it is, decrypts the other part of the message and calculates $(V^{q_{0,N}}_{0,N})^{p_{N,0}}$ from the received value $V^{q_{0,N}}_{0,N}$, and then extracts the beginning of the interval \emph{RI} from the received message to send the same size of interval (starting from \emph{RI}) from the calculated value, by then concatenates $ID_{N}$ and  $(V^{q_{0,N}}_{0,N})^{p_{N,0}}$, encrypts it with the shared session key and sends the message to the sink.

\item \emph{sink} $\rightarrow$ \emph{sensor node}: $K_{CS}$
    \newline
    In this step, the sink verifies the authenticity of the node as follows: if the interval of bits received in the message after decrypting is equal to the interval calculated by the sink in step 2, which means that ($V^{p_{N,0}}_{0,N})^{q_{0,N}}$ = ($V^{q_{0,N}}_{0,N})^{p_{N,0}}$, then the sink sends the key $K_{CS}$ used to commit the 200 bits in the second step to the node \emph{N}.
    \newline
     Otherwise, the sink stops the authentication mechanism and rejects all the data coming from this sensor node, until it succeeds its authentication.

\item \emph{Sensor node} $\rightarrow$ \emph{sink}: E($K_{0,N}$[$ID_{N}$ \textbar\textbar DATA])
    \newline
    If the authentication of the node \emph{N} is successfully done in step 4, the node receives the key commitment scheme $K_{CS}$ from the sink, which will enable it to decrypt the interval value of ($V^{p_{N,0}}_{0,N})^{q_{0,N}}$, and checks the authenticity of the sink.
    The node \emph{N} encrypts thereafter the DATA and the $ID_{N}$ and sends the message to the sink.
    \newline
    Otherwise the node \emph{N} denies the sink \emph{S} and sends no data.

    \end{enumerate}

\section{Security and efficiency Analysis}
\label{sec:Security_and_efficiency_Analysis}
In this section we discuss the performance of our solution in term of security, communication and computational cost efficiency.
As mentioned previously an adversary may initiate only external attacks by using computationally powerful devices such as personal computers. For example he/she can eavesdrop all the traffic between the different nodes and the sink, inject arbitrary messages, replay old ones, and spoof node identities.
\indent
It is also necessary to mention that an external adversary can launch denial of service (DoS) attacks, such as the black-hole attack, in which the attacker discards all received data, (these security attacks type is out from the scope of this paper). We make no assumption about the number of adversaries or their localizations.

\subsection{Security and Privacy Analysis}
We present in the following, the attacks that can be countered by our solution.

\paragraph{Forge node} In this attack the attacker acts as a legitimate node which can result an additional consumption of energy, not only of the sink but of the entire network since the used communication is a multi-hop broadcast, leading after that an attacker to inject false data. In our solution, before sending the data, the node must be authenticated to the sink. If the challenge imposed by the sink is not successfully done by the node, the sink will ignore all data coming from the node.
\paragraph{Forge sink} In this attack, the attacker acts as a legitimate sink to collect the pertinent data coming from different nodes. As our authentication scheme is mutual, the node, must be sure of the identity of the sink before sending any data. In addition, the data sent by the nodes are encoded with a key shared only between the legitimate sink and the relevant node.
\paragraph{Replay attack} In this attack, the attacker tries to maliciously or fraudulently replay the $(V^{p_{N,0}}_{0,N})^{q_{0,N}}$ interval values to make the sink think that it is one of the legitimate nodes in order to gain admission to the network, which can easily overrule the authentication mechanism. To prevent this attack we use the principle of commitment scheme that allows the sink to commit a message and reveal it later, which allows us to avoid this attack and also prevent the data injection attack that may result by making a successfully replay the $(V^{p_{N,0}}_{0,N})^{q_{0,N}}$ interval values.
\paragraph{Injection attack} As previously mentioned, this attack can be introduced after passing the replay attack. In this attack the attacker will try to inject false data into the network. The  main goal of this attack can be to circulate false information, to consume the resources of a node, or just saturate (overload) the network, it can also cause a bad decision that can have catastrophic consequences, especially when it comes to life or death of a human being.
\paragraph{Man in the Middle Attack} In this attack the main goal of the attacker is to get in between the legitimate node and the sink to control the entire conversation by establishing an independent connection with both of them, in order to sniff and intercept messages and by then trying to recover the secret or to gain access to sensitive information and perform malicious activities, or simply to get the pertinent data sent to the sink. However, in our solution, no information about the secret is disclosed, also the data sent to the sink are encrypted and no information on the key is sent over the communication channel.
\paragraph{Guessing Attack} In this attack the attacker tries to guess the key or the secret information by collecting several messages exchanged between the different nodes and the sink. Our proposed authentication protocol is effectively resisting to this attack since there is no secret information transmitted in BANZKP scheme. Even if in our scheme the Commitment Scheme key ($K_{CS}$) is sent in plain text, this latter gets changed with every communication and only 200 random interval is sent, thereby rendering the task of guessing shared values very difficult.
Moreover, the nodes also generates a random values (p and q) with every communication. Consequently the values also change randomly.
\paragraph{Attack on privacy} Privacy preservation of sensitive data in Body Area Networks is particularly a difficult challenge. One of the most common and easiest form of attack on data privacy is eavesdropping and passive monitoring. If the messages are not protected the attacker can easily understand and guess the disease that the patient suffer from. In our solution the messages are protected by cryptographic mechanism.

\subsection{Efficiency Analysis}
\label{sec:analysis}
In this subsection we  compare the communication and computational requirement of our protocol with respect to TinyZKP ~\cite{2}.
\paragraph*{Communication cost Analysis}The communication cost of our authentication scheme  can be achieved by four messages exchange and evaluated as follows:
\newline
 $2*L(V^{p/q})+ 2*L((V^{q_{0,n}}_{0,n})^{p_{n,0}})+ L(K_{CS}) $ = 1000 bits.  TinyZKP communication cost is at least:
 $L(M_{chall})+ L(ECDSA(M_{chall}))+ L(SHA-1(X_{m}))+ L(Y_{m})$ = 1710 bits.

\paragraph*{Computational cost Analysis}
Since in our solution the different keys are pre-distributed, the computational cost (in term of keys generation) is hence equal to zero. 
According to the literature~\cite{31}, the average number of modular multiplications for generating or verifying the identity is T*(k+2)/2, where T is the number of times we recalculate the modular multiplication, and k is the number of times we calculate a modular multiplication. In TinyZKP the authors use the modular multiplications to calculate the public keys. Therefore, the computational cost is 1*(20+2)/2=11, which requires not only additional computational resources but also a large memory in each node, especially for the sink node, that should hold the different public keys of each node (i.e. 120 public key for a 6 node network).

\section{Simulation settings and performance results}
\subsection{Simulation settings}
In this section, we evaluate our authentication and communication sending scheme by implementing it as an add-on to the convergecast routing protocol through the MiXiM project~\cite{26}, that joins and extends several existing simulation frameworks developed for wireless and mobile simulations in Omnet++~\cite{27}.

Our WBAN uses a ZigBee technology and consists of 7 sensor nodes deployed in a compact spatial region (in/on or around a human body). The sensor node that acts as the sink is the one deployed on the chest. The rest of the sensor nodes send a challenge/response messages with the sink until the approval of the identity of each one.

The sensor nodes, on which we have implemented our proposed protocol, have the following characteristics: 2.4\,GHZ, 3.3\,V Voltage, and the current draw is 10\,mAh

The performance of our protocol in terms of energy and memory consumption are evaluated by simulation and compared to the one achieved by TinyZKP which is to the best knowledge the only ZKP-based scheme defined for WBAN.

\subsection{Performance results}
\subsubsection{Energy Consumption}
\begin{figure}[!h]
\centering
\includegraphics[width=.9\columnwidth]{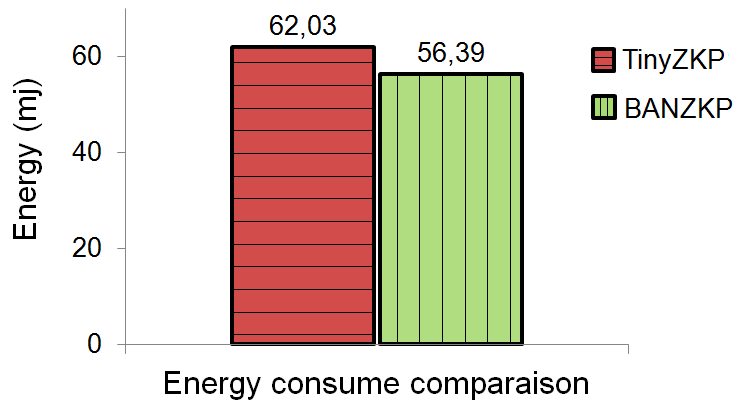}
\caption{Energy consume comparaison}
\label{energyCom1}
\end{figure}

As shown in Figure~\ref{energyCom1}. Our authentication scheme consumes less energy compared to TinyZKP since in our proposed protocol we used the Commitment Scheme that requires 1000 times less computational resources than ECDSA~\cite{11} and hence induces a lower energy consumption.
Additionally, in TinyZKP, the authors used a multiplicative modular operation to generate the public keys which also consumes energy, in contrast of our solution that uses the a pre-distributed keys. Furthermore, even if the number of data exchanges in TinyZKP is lower than in BANZKP, the communication cost has an important impact in terms of energy consumption and also in this case our proposed protocol consumes less energy than TinyZKP.

\subsubsection{Memory Consumption}
\begin{figure}[!h]
\centering
\includegraphics[width=.9\columnwidth]{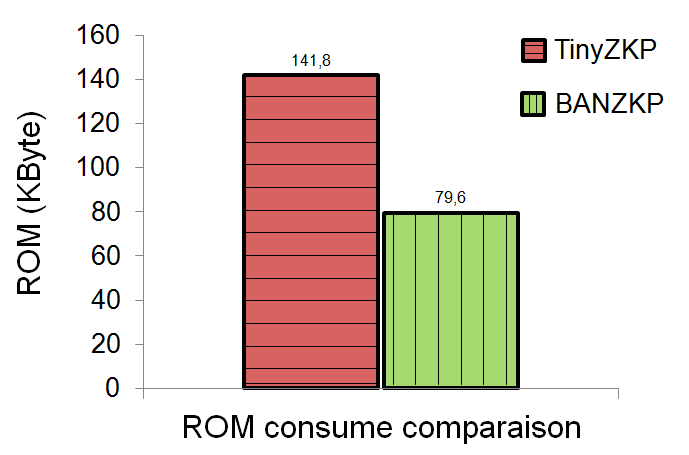}
\caption{Memory Consumption}
\label{MemoryComp}
\end{figure}

The required memory of the TinyZKP and BANZKP authentication protocols is given in Figure\ref{MemoryComp}. 
BANZKP that consumes  56.13\,\% less memory than TinyZKP.  In TinyZKP a big number of keys must be held in each node (20 public key and 20 private keys), especially in the sink node that must hold 120 public keys (in case of 6 nodes), plus 6 session keys for the authentication phase and 6 other keys for the data transmission. Furthermore the ECDSA and SHA-1 signature and verifications require additional memory resources. In contrast, our protocol uses a Commitment Scheme instead of ECDSA algorithm, and mades a simple comparison to verify the identity of the second party. Additionally, the number of keys used in our protocol is much lower than in TinyZKP.

\section{Conclusions}
\label{sec:conclusion}
In this paper
we propose and analyze the efficiency of a new lightweight authentication scheme for WBAN, BANZKP, which allows two nodes to make sure about the identity of each other and hence establish a trust relationship among the WBAN sensors to protect the subsequent wireless multi-hop communication throughout a low computational and memory requirement.  BANZKP is implemented as an an add-on to the convergecast routing protocol through the MiXiM project with the Omnet++ simulator. We then evaluated our protocol in terms of security and privacy as well as in terms of efficiency. The analysis shows that our protocol effectively resists to a variety of security and privacy attacks such as the replay attack and data injection attack.
%
BANZKP outperforms in terms of  energy and memory cost Tiny ZKP~\cite{2}  which is, to the best of our knowledge, the only ZKP scheme defined for WBAN. Our simulation results show that our authentication scheme BANZKP requires 56\% less memory and 10\% less energy compared to TinyZKP.

\bibliographystyle{abbrv}
\bibliography{PAPER}

\end{document}